\documentclass[aps,pra,amsmath,amssymb,amsfonts,superscriptaddress,lengthcheck,twocolumn,longbibliography]{revtex4-2}

\usepackage{dcolumn}    
\usepackage{graphicx}
\usepackage{amsmath}
\usepackage{bbm}
\usepackage{latexsym}
\usepackage{amsfonts}   
\usepackage{amssymb}
\usepackage{array}      
\usepackage{epsfig}
\usepackage{txfonts}
\usepackage{xcolor,braket}
\usepackage{ulem}
\usepackage{soul}
\usepackage{dsfont}
\usepackage{physics}
\usepackage{bm}
\usepackage{tensor}

\usepackage{relsize}

\usepackage{mathrsfs}

\usepackage{cancel}
\usepackage{enumitem}  

\usepackage[colorlinks=true,linkcolor=blue,urlcolor=blue,citecolor=blue,pdfusetitle]{hyperref}

\newcommand{\J}[1]{\hat{J}}
\newcommand{\vect}[1]{\mathbf{#1}}
\newcommand{\uvect}[1]{\underline{\mathbf{#1}}}

\begin{document}
%\title{Fast and Slow Dynamics in Multiparameter Quantum Estimation} 
\title{Geometric obstructions to quadratic time scaling in multiparameter quantum estimation}
\author{Eoin O'Connor}
\affiliation{Dipartimento di Fisica dell'Universit\`a di Milano, I-20133 Milan, Italy}
\author{Jiayu He}
\affiliation{Department of Physics, University of Helsinki, FI-00014 Helsinki, Finland}
\author{Matteo G. A. Paris}
\affiliation{Dipartimento di Fisica dell'Universit\`a di Milano, I-20133 Milan, Italy}
\author{Marco G. Genoni}
\affiliation{Dipartimento di Fisica dell'Universit\`a di Milano, I-20133 Milan, Italy}
\begin{abstract}
Unitary encoding of a single parameter provides quadratic enhancement in precision, with the quantum Fisher information scaling quadratically with the encoding time. However, when estimating multiple parameters simultaneously, this fundamental scaling is not guaranteed. Here, we establish a universal geometric obstruction that dictates when multiparameter quantum metrology fails to achieve simultaneous $t^{-2}$ scaling. By decomposing the Hamiltonian derivatives into components that commute and do not commute with the system Hamiltonian, we prove that linear dependence among the commuting components inevitably generates a {\it slow} parameter direction whose Fisher information remains bounded as O$(t^0)$, limiting the overall estimation precision. 
We demonstrate this mechanism in both discrete- and continuous-variable setups, including collective spin magnetometry and a generalized quantum harmonic oscillator, and contrast it with the Lipkin--Meshkov--Glick model where $t^{-2}$ decay is preserved.
%\sout{
%%This obstruction manifests across diverse physical systems: we demonstrate it explicitly in collective %spin magnetometry, generalized quantum harmonic oscillators, while we shot how in the Lipkin-Meshkov-Glick %model the $t^{-2}$ decay is preserved.
Remarkably, while the slow direction fundamentally limits the achievable precision, the measurement incompatibility between fast and slow directions decays as $1/t$, rendering the symmetric logarithmic derivative bound asymptotically saturable. Our framework provides a readily computable diagnostic, given by the Gram matrix of the diagonal generators, for identifying such obstructions in arbitrary multiparameter estimation problems. We further show that the bottleneck can be circumvented by relegating slow directions to nuisance parameters or by employing adaptive quantum control. 
%\sout{Our results establish fundamental limits on simultaneous multiparameter estimation and provide practical guidance for designing optimal quantum sensing protocols.} 
\end{abstract}
\maketitle
\section{Introduction}
\label{sec:intro}
The precise estimation of physical parameters governing quantum dynamics is the central goal of quantum metrology~\cite{GiovannettiPRL2006,paris2009,giovannetti2011a,hayashi2017}. For a well-isolated quantum system evolving under a time-independent Hamiltonian $H(g)$, the unitary dynamics $U_g(t) = e^{-i t H(g)}$ encodes information about the unknown parameter $g$ into the quantum state. In this single-parameter regime, the information gain is quantified by the quantum Fisher information (QFI)~\cite{helstrom1969,braunstein1994,holevo2011}, which typically scales quadratically with the total evolution time $t$~\cite{GiovannettiPRL2006,pang2014,RafalTimeScaling2025}. This translates to an estimation variance that scales as $\mathcal{O}(t^{-2})$.

However, many practical and fundamental sensing tasks require the simultaneous estimation of multiple parameters. In multiparameter quantum estimation, the situation becomes significantly more intricate~\cite{liu2019,albarelli2020a,demkowicz2020a}. The ultimate precision is no longer dictated solely by the temporal scaling of individual parameters, but is also restricted by the underlying geometry of the multiparameter landscape. Phenomena such as measurement incompatibility~\cite{Genoni2013,VidrighinNatCommm,ragy2016,Roccia_2018,AlbarelliHolevo2019,Carollo_2019,Razavian2020,Candeloro_2021,Conlon2021,albarelli2022,candeloro2024,he2025} (arising from the non-commutativity of optimal observables for different parameters), parameter sloppiness~\cite{gessner2018,frigerio2025,sharma2025,he2025a}, and singularities in the quantum Fisher information matrix (QFIM)~\cite{goldberg2021,yang2025,mihailescu2025a} can severely degrade the achievable precision. 

In this work, we investigate the dynamical conditions under which multiparameter estimation protocols fail to achieve quadratic time scaling for parameters encoded in the time-independent Hamiltonian of a quantum probe. By decomposing the Hamiltonian derivatives into components that commute (diagonal) and do not commute (off-diagonal) with the system Hamiltonian, we reveal a universal geometric obstruction. Specifically, we demonstrate that when the diagonal components of the Hamiltonian derivatives are linearly dependent, the parameter space inevitably develops a ``slow'' direction. Along this direction, the estimation variance is fundamentally bottlenecked to an $\mathcal{O}(t^0)$ scaling, thus destroying simultaneous $\mathcal{O}(t^{-2})$ scaling. 
While isolated instances of this behavior have been observed in specific models~\cite{HaidongPRL2016,hou2020}, our framework provides a rigorous generalization that unveils the underlying geometric mechanism. We illustrate these results through paradigmatic continuous- and discrete-variable models, and discuss how this $t^0$ bottleneck impacts measurement incompatibility, the treatment of nuisance parameters, and strategies for optimal quantum control. Additionally, using our formalism, we identify the optimal input probe states that achieve the ultimate bound on the scaling.

The manuscript is organized as follows: In Sec.~\ref{sec:multiparameter_qmetro} we provide a basic introduction to multiparameter quantum estimation theory. In Sec.~\ref{sec:scaling_proof} we establish the general framework and derive the conditions under which quadratic time scaling of the estimation precision is lost. In Secs.~\ref{sec:collectivespin}, ~\ref{sec:QHO} and ~\ref{sec:LMG} we apply the framework above to three paradigmatic examples of multiparameter quantum estimation, while in Sec.~\ref{sec:discussion} we discuss the relationship of our general results to optimal quantum control strategies, measurement incompatibility, and the presence of nuisance parameters. We conclude the manuscript in Sec.~\ref{sec:conclusions} with some final remarks and outlook.
\section{Multiparameter quantum metrology}
\label{sec:multiparameter_qmetro}
Multiparameter quantum metrology focuses on the ultimate limits on the simultaneous estimation of $p$ unknown parameters $\vect{g} = (g_1,\dots,g_p)^T$ that characterize a family of quantum states $\hat\rho_{\vect{g}}$. To estimate these unknown parameters, one performs a general quantum measurement characterized by a positive operator-valued measure (POVM). According to Born’s rule, the probability of obtaining a specific measurement outcome $x$ is $p(x|\vect{g} ) = \Tr(\hat{\rho}_{\vect{g} }\hat{\Pi}_{x})$, where $\hat{\Pi}_{x}$ denotes the corresponding POVM element. Based on the observed outcome $x$, a locally unbiased estimator function $\check{\vect{g}}(x)$ is employed to infer the parameter values. The performance of this estimator is quantified by the mean-squared-error matrix:
\begin{align}
    \Sigma_{\vect{g} }(\check{\vect{g}})=\int \! dx \, p(x|\vect{g} ) \left(\check{\vect{g}}(x)-\vect{g} \right) \left(\check{\vect{g}}(x)-\vect{g} \right)^{T}.
\end{align}
Given a real positive semidefinite matrix $W$, one can derive a scalar bound on the overall precision in the estimation as 
\begin{align}
\Tr \left[ W \Sigma_{\vect{g} }(\check{\vect{g}}) \right] \geq \Tr \left[ W F^{-1} \right],
\label{eq:SLD_Matrix_CRB}
\end{align}
where $F$ is the quantum Fisher information matrix (QFIM) with elements
\begin{equation}
    F_{\mu\nu} = \tfrac{1}{2}\,\mathrm{Tr}\!\left[\hat\rho_{\bf g}\{\hat{L}_\mu,\hat{L}_\nu\}\right],
\end{equation}
where $\hat{L}_\mu$ are the symmetric logarithmic derivative (SLD) operators defined by $\partial_\mu \hat{\rho}_{\bf g} = \tfrac{1}{2}(\hat{L}_\mu \hat{\rho}_{\bf g} + \hat\rho_{\bf g} \hat{L}_\mu)$, and we define the derivative with respect to a specific parameter $\partial_\mu = \partial/\partial g_\mu$. In the following, we will mainly focus on the case $W=\mathbb{I}$, where one obtains the bound on the sum of the variances for each parameter estimator as
\begin{equation}
    \label{eq:SLD_bound}
    \sum_{i = 1}^p \delta\check{g}_i^2 \geq \Tr(F^{-1}).
\end{equation}
We refer to this as the SLD bound.

We will focus on pure state families, generated by a unitary encoding, $|\psi_{\bf g}\rangle =\hat{U}_{\bf g} |\psi_0\rangle$ with $\hat{U}_{\bf g} = e^{-i t \hat{H}({\bf g})}$, where the dependence on the parameters is encoded in the Hamiltonian operator $\hat{H}({\bf g})$ (note that in the rest of the manuscript we will often omit the dependence of the Hamiltonian on the parameter vector ${\bf g}$); in this case the QFIM can be written in terms of the covariance matrix of the local generators $\hat{\mathcal{H}}_\mu = i(\partial_\mu \hat U^\dagger)\hat U$, yielding~\cite{boixo2007a,liu2015a}
\begin{equation}
    \label{eq:qfim_def}
    F_{\mu\nu} = 4\,\mathrm{Cov}_{\ket{\psi_0}}(\hat{\mathcal H}_\mu, \hat{\mathcal H}_\nu),
\end{equation}
where we have defined $\mathrm{Cov}_{\ket{\psi_0}}(\hat{A}, \hat{B}) = \langle \psi_0 | \{\hat{A},\hat{B}\}|\psi_0\rangle/2 - \langle \psi_0 |\hat{A} |\psi_0\rangle\langle\psi_0 | \hat{B} | \psi_0\rangle$, with $\{\hat{A},\hat{B}\}$ denoting the anti-commutator.

Unlike the single-parameter case, the SLD bound is not always saturable due to measurement incompatibility, which arises when the optimal observables for different parameters do not commute. While one can derive alternative bounds like the RLD bound, or tighter bounds like the Holevo bound~\cite{liu2019,albarelli2020a,demkowicz2020a}, the SLD bound alone is sufficient for our objectives. Specifically, since we aim to prove a no-go theorem on the achievable time scaling in multiparameter quantum metrology, analyzing the SLD bound is sufficient.
\section{When does quadratic time scaling fail?}
\label{sec:scaling_proof}
In this section, we derive a general condition under which multiparameter estimation fails to achieve quadratic time scaling ($t^{-2}$) of the SLD quantum Cram\'er--Rao bound. 

The first scenario is a singular QFIM. Simply, if the QFIM is singular, then it cannot be inverted, and it is impossible to estimate the unknown parameters simultaneously. This occurs when the local generators $\{\hat{\mathcal{H}}_\mu\}$ are linearly dependent.

We are interested in scenarios where the QFIM is non-singular yet, nonetheless, quadratic time scaling fails. We start by decomposing the Hamiltonian derivative $\partial_\mu \hat{H}$ into a part $\hat{D}_\mu$ which commutes with $\hat{H}$, and a part $\hat{O}_\mu$ which does not commute:
\begin{equation}
    \partial_\mu \hat{H} = \hat{D}_\mu + \hat{O}_\mu, \quad \text{where } [\hat{H}, \hat{D}_\mu] = 0.
\end{equation}
For finite-dimensional systems, we can write the Hamiltonian in the form $\hat{H} = \sum_\alpha E_\alpha \hat{P}_\alpha$ where $\hat{P}_\alpha$ is the projector onto the (potentially degenerate) subspace with eigenvalue $E_\alpha$. In this basis, we have
\begin{align}
    \hat{D}_\mu &= \sum_\alpha \hat{P}_\alpha (\partial_\mu \hat{H}) \hat{P}_\alpha, \\
    \hat{O}_\mu &= \sum_{\alpha \neq \beta} \hat{P}_\alpha (\partial_\mu \hat{H}) \hat{P}_\beta\,,
\end{align}
showing how $\hat{D}_\mu$ is (block-)diagonal, while $\hat{O}_\mu$ is off-(block-)diagonal.
Substituting these formulas into the integral expression for the generator, we obtain:
\begin{align}
    \hat{\mathcal{H}}_\mu(t) &= -\int_0^t ds \, e^{i\hat{H}s} (\hat{D}_\mu + \hat{O}_\mu) e^{-i\hat{H}s} \nonumber \\
    &= -t \hat{D}_\mu - \hat{K}_\mu(t). \\
    \hat{K}_\mu(t) :\!\!&= \sum_{\alpha \neq \beta} \frac{e^{i(E_\alpha - E_\beta)t} - 1}{i (E_\alpha - E_\beta)} \hat{P}_\alpha (\partial_\mu \hat{H}) \hat{P}_\beta \label{eq:off_diag_local}
\end{align}
Notice that $\hat{K}_\mu(t) = \mathcal{O}(t^0)$ as $t \to \infty$. We remark that similar decompositions have been noted before~\cite{pang2014, sidhu2018, sidhu2020}. Since the QFIM is invariant under the transformation $\hat{\mathcal{H}}_\mu \to \hat{\mathcal{H}}_\mu + c_\mu \hat{\mathbbm{1}}$, the correct quantity to consider is the traceless diagonal generator 
\begin{equation}
\hat{\mathcal{D}}_\mu = \hat{D}_\mu - \Tr[\hat{D}_\mu]\frac{\hat{\mathbbm{1}}}{d}.
\end{equation}
When the Hamiltonian is non-degenerate, the decomposition has a particularly clear interpretation; $t \hat{D}_\mu$ and $\hat{K}_\mu(t)$ quantify the information respectively in the eigenvalues and in the eigenvectors~\cite{pang2014}. It becomes clear that if $\hat{\mathcal{D}}_\mu = 0$ for any parameter, then $\delta\check{g}_\mu^2 \geq \mathcal{O}(t^0)$ and quadratic time scaling is lost for the SLD bound.

The final scenario we consider is that in which $\hat{\mathcal{D}}_\mu \neq 0$ for all parameters and the QFIM is non-singular. This means that all parameters can individually be estimated with quadratic-in-time scaling. Even in this case, we can lose quadratic time scaling in multiparameter estimation. The condition under which it fails, and we fall back to $t^0$ scaling, is when $\{\hat{\mathcal{D}}_\mu\}$ are linearly dependent. We know that $\text{Tr}(F^{-1}) = \sum_k \lambda_k^{-1}$, where $\lambda_k$ are the eigenvalues of $F$. The asymptotic scaling of the error is dominated by the smallest eigenvalue $\lambda_{\min}$. For any symmetric matrix, the smallest eigenvalue is given by
\begin{equation}
    \lambda_{\min} = \min_{\uvect{u}} \left( \uvect{u}^T F \uvect{u} \right) = \min_{\uvect{u}} 4 \text{Var}_{\psi_0} \left( \sum_\mu u_\mu \hat{\mathcal{H}}_\mu(t) \right),
    \label{eq:min_eval}
\end{equation}
where we have introduced the notation $\uvect{u}$ to denote unit vectors and $\text{Var}_{\psi_0} (\hat{A}) = \langle\psi_0 | \hat{A}^2 |\psi_0\rangle -  \langle\psi_0 | \hat{A} |\psi_0\rangle^2 $.
We now assume that the set of traceless diagonal operators $\{ \mathcal{D}_\mu \}$ is linearly dependent.
%and prove that the minimum eigenvalue scales as $\mathcal{O}(t^0)$. 
Linear dependence implies that there exists a unit vector $\uvect{w}$ such that:
\begin{equation} \label{eq:null_condition}
    \sum_\mu w_\mu \hat{\mathcal{D}}_\mu = 0.
\end{equation}
Therefore,
\begin{align}
    \sum_\mu w_\mu \hat{\mathcal{H}}_\mu(t) = -t \underbrace{\left( \sum_\mu w_\mu \hat{D}_\mu \right)}_{\propto\hat{\mathbbm{1}}} - \sum_\mu w_\mu \hat{K}_\mu(t).
\end{align}
Consequently, the variance along the direction $\uvect{w}$ depends only on the bounded operators $\hat{K}_\mu$:
\begin{equation}
    \uvect{w}^T F \uvect{w} = 4 \text{Var}_{\psi_0} \left( \sum_\mu w_\mu \hat{K}_\mu(t) \right).
\end{equation}
Since all $\hat{K}_\mu(t) \sim t^0$, the variance minimized over all possible vectors $\uvect{u}$, and thus the smallest eigenvalue in Eq.~\eqref{eq:min_eval}, is also $\mathcal{O}(t^0)$. Therefore,
\begin{equation}
    \text{Tr}(F^{-1}) \ge \frac{1}{\lambda_{\min}} \sim t^0.
    \label{eq:degenerate_scaling}
\end{equation}
This proof represents the main result of the paper. Since it is based on the properties of the generators, it holds for all initial states. Notably, while ancilla-assisted strategies are often used in quantum metrology to improve precision limits, in particular in the presence of noise~\cite{RafalLorenzo2014,Huang2018,Sbroscia2018}, and more recently also in the multiparameter scenario~\cite{hou2020, hou2021, song2024agnostic}, they cannot bypass this specific bottleneck. If the probe is entangled with an unparameterized ancilla $A$, the local generators become $\hat{\mathcal{H}}_\mu(t) \otimes \hat{I}_A$; the linear dependence of the diagonal components is preserved, and the scaling of the variance remains strictly limited to $O(t^0)$.

Numerically, linear dependence can be checked by constructing the Gram matrix
\begin{equation}
    G_{\mu \nu} = \Tr[\hat{\mathcal{D}}_\mu \hat{\mathcal{D}}_\nu].
    \label{eq:Gram}
\end{equation}
When $\{\mathcal{D}_\mu\}$ are linearly dependent, then $\det G = 0$. We can thus identify the determinant of the Gram matrix $\det G$ as a readily computable quantity that directly allows us to verify whether our multiparameter estimation problem is restricted to an $O(t^0)$ scaling.
More information can also be obtained by directly checking the eigenvalues of the matrix $G$: in fact, the number of zero eigenvalues indicates exactly how many redundant operators one has. The non-zero eigenspace identifies the parameter directions whose diagonal generators are nonzero and hence can, in principle, support quadratic scaling.\\

An immediate and powerful consequence of this geometric obstruction arises when considering finite-dimensional systems. Consider a $d$-dimensional, non-degenerate Hamiltonian $H = \sum_{n=1}^d E_n \ketbra{n}$. By the Hellmann--Feynman theorem, the commuting component of the Hamiltonian derivative is
\begin{equation}
    \hat{D}_\mu = \sum_{n=1}^d (\partial_\mu E_n) |n\rangle\langle n|.
\end{equation}
As discussed above, only the traceless part of this operator is relevant for the QFIM, since an identity component contributes only a parameter-dependent global phase. Therefore, each $\hat{D}_\mu$ can be represented as a vector in a $d-1$ dimensional space. It follows that if we try to estimate $p \geq d$ parameters, $\{\hat{\mathcal{D}}_\mu\}$ is guaranteed to be linearly dependent. This imposes a fundamental dimensional bound on multiparameter quantum metrology: in a $d$-dimensional quantum system, at most $d-1$ parameters can be simultaneously estimated with quadratic time scaling $O(t^{-2})$. Any attempt to simultaneously estimate $d$ or more parameters will inevitably force at least one parameter direction into the $O(t^0)$ precision bottleneck.

If an eigenvalue has degeneracy $r_\alpha$, the commuting block-diagonal Hermitian operators span a real vector space of dimension $r_\alpha^2$. Therefore, in general, the maximum number of parameters that can be estimated with $O(t^{-2})$ scaling is $\sum_i r_i^2 -1$. 

In the following, we will apply this framework to three different examples: i) parameter estimation in spin systems whose Hamiltonians are linear in the collective spin operators, which encompasses multiparameter quantum magnetometry; ii) parameter estimation of a quantum harmonic oscillator Hamiltonian; and iii) parameter estimation in the Lipkin-Meshkov-Glick model.
\section{Example 1: Collective spin system}
\label{sec:collectivespin}
We consider a system of $N$ spin-$1/2$ particles whose evolution is governed by a time-independent Hamiltonian
\begin{equation}
    \hat{H} = \vect b(g,h)\cdot \hat{\bf J},
\end{equation}
where $\hat{\bf J} = (\hat{J}_x, \hat{J}_y, \hat{J}_z)$ is the vector of collective spin operators. Variations of this problem have been studied in ~\cite{kolenderski2008, jing2015, baumgratz2016a, yang2022multiparameter} and the ultimate precision limits are known~\cite{hou2020} for specific parameter encodings. Here we will present a simple derivation of the optimal bound for the two-parameter case in the long-time limit; importantly, this bound is valid for arbitrary parameter encodings. It is helpful to reparameterize the Hamiltonian in terms of a frequency, $\Omega$ and a unit vector $\uvect{n}$ that determines the direction
\begin{equation}
    \hat{H} = \Omega(g,h)\,\left(\uvect n(g,h)\cdot \hat{\bf J} \right).
\end{equation}

\subsection{Diagonal/off-diagonal decomposition.}

In Appendix~\ref{app:generator_collective}, we show that the local generator can be written in the form
\begin{equation}
    \hat{\mathcal{H}}_\mu = \vect{h}_\mu \cdot \hat{\vect{J}},
\end{equation}
where the vector $\vect{h}_\mu$ is given by
\begin{equation}
    \vect{h}_\mu = -t(\partial_\mu\Omega)\,\uvect n + \vect{h}_\mu^\perp(t),
    \label{eq:h_vector2}
\end{equation}
with
\begin{equation}
    \vect{h}_\mu^\perp(t) = - \sin(\Omega t)\,\partial_\mu\uvect n + (1-\cos(\Omega t))\,(\uvect n\times\partial_\mu\uvect n) \,.
    \label{eq:perp_vec}
\end{equation}
The part of $\vect{h}_\mu$ that ``commutes'' with the Hamiltonian is the part parallel to $\uvect n(g,h)$, while $\partial_\mu\uvect n$ and $\uvect n\times\partial_\mu\uvect n$ are both perpendicular. Therefore we have the decomposition
\begin{equation}
\hat{D}_\mu=\partial_\mu\Omega\left(\uvect n \cdot \hat{\vect{J}} \right),
\qquad
\hat{K}_\mu(t)=-\vect{h}_\mu^\perp(t) \cdot \hat{\vect{J}}.
\end{equation}
It becomes clear that the set $\{\hat{D}_g,\hat{D}_h\}$ is always linearly dependent: both are proportional to $\uvect n \cdot \hat{\vect{J}}$.

The QFIM entries in Eq.~\eqref{eq:qfim_def} take the form
\begin{equation}
    F_{\mu \nu} = 4 \vect{h}_\mu^T \, \Gamma \, \vect{h}_\nu
    \label{eq:cov_decomp}
\end{equation}
where $\Gamma$ is the covariance matrix of the collective spin operators with respect to the initial state
\begin{equation}
    \Gamma_{ij} = \frac12\langle\{\Delta \hat{J}_i,\Delta \hat{J}_j\}\rangle_{\psi_0}
    \label{eq:cov_mat}
\end{equation}

\subsection{Slow direction in parameter space}
\label{sec:slow_coll}
The leading $t^2$ contribution to the QFIM is
\begin{equation}
    4 t^2\,\mathrm{Cov}_{\ket{\psi_0}}(\hat{D}_\mu, \hat{D}_\nu) = 4t^2\mathrm{Var}_{\ket{\psi_0}}\left(\uvect n \cdot \hat{\vect{J}} \right)\nabla\Omega\nabla\Omega^T.
\end{equation}
Therefore, the fast direction in parameter space is $\nabla\Omega =( \partial_g \Omega , \partial_h \Omega )^T $. The slow direction is orthogonal, given by the vector,
\begin{equation}
    \vect{w} = \begin{pmatrix} \partial_h \Omega \\ -\partial_g \Omega \end{pmatrix}.
\end{equation}
The squared norm of this vector is $\|\vect{w}\|^2 = (\partial_g \Omega)^2 + (\partial_h \Omega)^2$, leading to $\uvect{w} =\vect{w} /\|\vect{w}\|$.
We transform the local generators into this fast and slow basis
\begin{equation}
    \hat{\mathcal{H}}_f = -t \|\vect{w}\| \left(\uvect n \cdot \hat{\vect{J}}\right) + \mathcal{O}(t^0), \quad \hat{\mathcal{H}}_s = \frac{\vect{h}_{\vect{w}} \cdot \hat{\vect{J}}}{\|\vect{w}\|} 
\end{equation}
where we have defined the vector
\begin{align}
    \vect{h}_{\vect{w}} &= \partial_h \Omega \vect{h}_g - \partial_g \Omega \vect{h}_h \nonumber \\
    &=  -\sin(\Omega t)\vect{D} + (1-\cos(\Omega t)) (\uvect{n}\times\vect{D}),
\end{align}
with coupling vector $\vect{D} := (\partial_h \Omega)\partial_g\uvect n - (\partial_g \Omega)\partial_h\uvect n$.

The leading order of the QFIM entries are
\begin{align}
    F_{ff} &= 4 t^2 \|\vect{w}\|^2 \mathrm{Var}_{\ket{\psi_0}}\left(\uvect n \cdot \hat{\vect{J}} \right) + \mathcal{O}(t) \\
    F_{fs} &= -4 t \mathrm{Cov}_{\ket{\psi_0}}\left(\uvect n \cdot \hat{\vect{J}}, \vect{h}_{\vect{w}} \cdot \hat{\vect{J}}\right) + \mathcal{O}(t^0) \\
    F_{ss} &= \frac{4}{\|\vect{w}\|^2} \mathrm{Var}_{\ket{\psi_0}}\left(\vect{h}_{\vect{w}} \cdot \hat{\vect{J}} \right)
\end{align}
Then the smallest eigenvalue takes the form
\begin{equation}
    \lambda_\text{min} = F_{ss} - \frac{F_{fs}^2}{F_{ff}} + \mathcal{O}(t^{-1}).
\end{equation}

\subsection{Optimal initial state}
To maximize $\lambda_\text{min}$, and therefore minimize $\Tr(F^{-1})$, we must maximize the variance in the direction $\vect{h}_{\vect{w}}$ while keeping $F_{fs}$ small. Luckily, in this case, we can do both simultaneously. For a system of $N$ spin-$1/2$ particles, the standard basis is the Dicke basis $\ket{j,m}$, which are the simultaneous eigenstates of the collective spin operator, $\hat{\vect{J}}^2$, and the $z$ angular momentum operator, $\hat{J}_z$. However, this choice of basis is arbitrary and we can define a new basis $\ket{j,m}_{\uvect{y}}$ which are the simultaneous eigenstates of the collective spin operator, $\hat{\vect{J}}^2$, and the angular momentum operator, $\uvect{y} \cdot \hat{\vect{J}}$. The state that maximizes $\text{Var}_{\vect{h}_{\vect{w}}}$ is a Greenberger–Horne–Zeilinger (GHZ) state in the $\uvect{h}_{\vect{w}}$ basis. This can be written in the form $\ket{\psi_0} = (\ket{j,j}_{\uvect{h}_{\vect{w}}} + e^{i \phi}\ket{j,-j}_{\uvect{h}_{\vect{w}}})/\sqrt{2}$ with $j = N/2$, which corresponds to a variance 
\begin{align}
\text{Var}_{\ket{\psi_0}}\left( \vect{h}_{\vect{w}} \cdot \hat{\vect{J}}\right) = \frac{\|\vect{h}_{\vect{w}}\|^2N^2}4
\end{align}
with
\begin{align}
    \|\vect{h}_{\vect{w}}\|^2 &= \sin^2(\Omega t) \|\vect{D}\|^2 + (1-\cos(\Omega t))^2 \|\uvect{n}\times\vect{D}\|^2 \nonumber \\
    &= 4\sin^2(\Omega t/2)\|\vect{D}\|^2.
    \label{eq:perp_vec_mag}
\end{align}
Importantly, the covariance vanishes for this state
\begin{equation}
    \mathrm{Cov}_{\ket{\psi_0}}\left(\uvect n \cdot \hat{\vect{J}}, \vect{h}_{\vect{w}} \cdot \hat{\vect{J}}\right) = 0.
\end{equation}
Combining these results, we have
\begin{equation}
    \label{eq:mag_QFI}
    \min_{\ket{\psi_0}} \Tr(F^{-1}) \simeq \frac{(\partial_g\Omega)^2+(\partial_h\Omega)^2}{4\sin^2(\Omega t/2)N^2\|\vect{D}\|^2}, \quad t\to\infty.
\end{equation}

\subsection{Comparison to previous results}
A very general result was derived for this problem in~\cite{hou2020}, where one has considered a multiparameter estimation problem defined by a unitary encoding $|\psi\rangle = \hat{U} |\psi_0\rangle$, where
\begin{align}
\hat{U} = e^{- 2i \alpha \uvect{n} \cdot \hat{\vect{J}}} \,
\end{align}
with $\uvect{n} = (\sin\theta\cos\phi,\sin\theta\sin\phi,\cos\theta)^T$.
Transforming $\alpha = \Omega t /2$ and taking the $t \to \infty$ limit their bound becomes
\begin{equation}
    w_{\Omega} \delta \check{\Omega}^2 + w_{\theta} \delta \check{\theta}^2 + w_{\phi} \delta \check{\phi}^2 \geq \frac{\left(\sqrt{w_{\theta}}+ \frac{\sqrt{w_{\phi}}}{|\sin \theta|}\right)^2}{4\sin^2(\Omega t/2)N(N+2)}.
    \label{eq:Hou}
\end{equation}
Where $(w_\Omega,w_\theta,w_\phi)$ can be thought of as the diagonal entries of a weight matrix. However, this diagonal-weight result cannot be applied directly to a generic reparameterization. Suppose we want to estimate an arbitrary parameter vector $\bm{\eta}$. What we want to calculate is a bound on $\langle \delta \bm{\check \eta}^T\delta \bm{\check \eta}\rangle$. Defining the parameter vector $\vect{x} = \{\Omega, \theta, \phi\}$, the transformation between $\bm{\eta}$ and $\vect{x}$ is given by a Jacobian $J$, which also determines the transformation of the estimator variances $\delta \bm{\check \eta} = J \delta \vect{\check x}$. Substituting this into the variance, we have
\begin{equation}
    \langle \delta \bm{\eta}^T\delta \bm{\eta}\rangle = \langle \delta \vect{x}^T (J^T\! J) \delta \vect{x}\rangle := \langle \delta \vect{x}^T W \delta \vect{x}\rangle.
\end{equation}
Therefore, a generic parameter transformation induces a non-diagonal weight matrix in the ($\Omega,\theta,\phi$) coordinates, and Eq.~\eqref{eq:Hou} cannot be applied directly. The fast/slow derivation leading to Eq.~\eqref{eq:mag_QFI} avoids this restriction by working directly with the arbitrary two-parameter encoding and optimizing the single slow generator selected by the direction orthogonal to $\nabla\Omega$.

When the induced weight matrix happens to be diagonal, the two approaches agree up to the factor $(N+2)/N$. This factor reflects the difference between the two optimization problems: Eq.~\eqref{eq:Hou} comes from a three-coordinate problem with two angular slow directions, for which the relevant collective-spin variance budget is $N(N+2)/4$, whereas Eq.~\eqref{eq:mag_QFI} concerns a two-parameter problem with only one slow direction, whose maximal variance is $N^2/4$.
\section{Example 2: The quantum harmonic oscillator}
\label{sec:QHO}
We consider the paradigmatic example of the unitary encoding $\hat{U} = e^{-i H(g,h) t}$, induced by a Hamiltonian quadratic in the position and momentum operators 
\begin{equation}
\hat H(g,h)=A(g,h) \frac{\hat p^2}{2}+B(g,h) \frac{\hat x^2}{2},
\qquad [\hat x,\hat p]=i,
\end{equation}
with $A>0$, $B>0$. We want to estimate two parameters that can be encoded in any way in $A$ and $B$. This problem encompasses several scenarios, ranging from the standard quantum harmonic oscillator Hamiltonian parametrized by its mass and frequency to more exotic cases, such as the effective Hamiltonian governing the low-energy spin sector of the quantum Rabi model in the large-qubit-frequency limit~\cite{mihailescu2025,mihailescu2026}.
It is convenient to reparameterize in terms of a frequency and a squeezing parameter,
\begin{equation}
\omega := \sqrt{AB},
\qquad
r := \frac14 \ln\!\left(\frac{B}{A}\right),
\end{equation}
so that $A=\omega e^{-2r}$ and $B=\omega e^{2r}$. Defining the annihilation operator and the corresponding number operator
\begin{equation}
\hat c=\frac{1}{\sqrt{2}}\left(e^{r}\hat x+i e^{-r}\hat p\right),\qquad \hat n=\hat c^\dagger \hat c,
\end{equation}
the Hamiltonian takes the standard form
\begin{equation}
\hat H=\omega\left(\hat n+\frac12\right).
\end{equation}

\subsection{Diagonal/off-diagonal decomposition}
The Hamiltonian derivative is
\begin{equation}
\partial_\mu \hat H
=
(\partial_\mu\omega)\left(\hat n+\frac12\right)
+
\omega(\partial_\mu r)\left(\hat c^2+\hat c^{\dagger 2}\right).
\end{equation}
This can be decomposed into the diagonal and off-diagonal components,
\begin{equation}
\hat{D}_\mu=(\partial_\mu\omega)\left(\hat n+\frac12\right),
\qquad
\hat{O}_\mu=\omega(\partial_\mu r)\left(\hat c^2+\hat c^{\dagger 2}\right).
\end{equation}
The diagonal components are once again clearly collinear. 

Using Eq.~\eqref{eq:off_diag_local} we obtain
\begin{equation}
\hat K_\mu(t) = \omega(\partial_\mu r)\!\left[
\frac{1-e^{-2i\omega t}}{2i\omega}\hat c^2
+
\frac{e^{2i\omega t}-1}{2i\omega}\hat c^{\dagger 2}
\right].
\end{equation}
The scalar prefactors are bounded in time, so $\hat K_\mu(t)\sim O(t^0)$, although $\hat K_\mu(t)$ still contains unbounded operators. Note that although this is guaranteed in systems with a finite Hilbert space dimension, it must be verified case by case in continuous variable systems.

\subsection{Slow direction in parameter space}
\label{sec:slow_qho}
In direct analogy with Sec.~\ref{sec:slow_coll}, the smallest eigenvalue must be in the direction perpendicular to the frequency gradient $\nabla \omega$
\begin{equation}
\vect{w}:=
\begin{pmatrix}
\partial_h\omega\\
-\partial_g\omega
\end{pmatrix},
\qquad
\uvect{w}=\frac{\vect{w}}{\|\vect{w}\|},
\qquad
\|\vect{w}\|^2=(\partial_g\omega)^2+(\partial_h\omega)^2.
\end{equation}
Indeed,
\begin{equation}
\sum_{\mu\in\{g,h\}} w_\mu \hat{D}_\mu = 0,
\end{equation}
and hence the corresponding linear combination of local generators has no $t$-term:
\begin{equation}
\sum_\mu w_\mu \hat H_\mu(t) = -\sum_\mu w_\mu \hat K_\mu(t)
=
-\mathcal{J}\sin(\omega t)\hat Q(t),
\end{equation}
with
\begin{align}
    \mathcal{J} &= (\partial_h\omega)(\partial_g r)-(\partial_g\omega)(\partial_h r) \\
    \hat Q(t) &= \left(e^{-i\omega t} \hat{c}^2 + e^{i \omega t} \hat{c}^{\dag 2}\right).
\end{align}
$\mathcal{J}$ is the Jacobian determinant of the parameter transformation from $\{\omega, r\}$ to $\{g,h\}$.

In the fast/slow basis, the QFIM has the asymptotic form
\begin{align}
    F_{ff} &= 4 t^2 \|\vect{w}\|^2 \mathrm{Var}_{\ket{\psi_0}}\left(\hat{n} \right) + \mathcal{O}(t) \\
    F_{fs} &= 4 t \mathcal{J} \sin(\omega t)\mathrm{Cov}_{\ket{\psi_0}}\left(\hat{n}, \hat{Q}(t)\right) + \mathcal{O}(t^0) \\
    F_{ss} &= \frac{4 \mathcal{J}^2 \sin^2(\omega t)}{\|\vect{w}\|^2} \mathrm{Var}_{\ket{\psi_0}}\left(\hat{Q}(t)\right)
\end{align}

\subsection{Optimal initial state}
Once again, our goal is to maximize
\begin{equation}
    \lambda_\text{min} = F_{ss} - \frac{F_{fs}^2}{F_{ff}} + \mathcal{O}(t^{-1}).
\end{equation}
In continuous-variable systems, the infinite-dimensional Hilbert space allows for operators with unbounded variances. To extract physically meaningful and experimentally relevant bounds, we restrict our analysis to Gaussian states with a fixed average of bosonic excitations $\langle\hat{n}\rangle$. Unfortunately, we cannot simultaneously maximize the variance of $\hat{Q}(t)$ while keeping the covariance equal to zero. In Appendix~\ref{app:qho_min_eig} we prove that the maximum value of the minimum eigenvalue is
\begin{equation}
    \lambda_\text{min}^{\mathrm{opt}}
    \approx
    \frac{4 \mathcal{J}^2\sin^2(\omega t)}{\|\vect{w}\|^2}
    \left[
    4\langle \hat{n} \rangle(\langle \hat{n} \rangle+1)+2
    \right].
    \label{eq:lmin_opt}
\end{equation}
This is achieved with a displaced squeezed vacuum state with a displacement parameter
\begin{equation}
    \alpha = \sqrt{
    \frac{\langle \hat{n} \rangle(\langle \hat{n} \rangle+1)}{2\langle \hat{n} \rangle+1}
    }
    e^{i\frac{\pi + 2 \omega t}{4}}
\end{equation}
and a squeezing parameter $\zeta = s e^{i \theta}$ with
\begin{equation}
    \sinh^2 s
    =
    \frac{\langle \hat{n} \rangle^2}{2\langle \hat{n} \rangle+1}, \quad \theta = \frac{\pi + 2 \omega t}{2}.
\end{equation}
The resulting optimized SLD bound is
\begin{equation}
    \min_{\ket{\psi_0}}
    \mathrm{Tr}\left(F^{-1}\right)
    \approx
    \frac{
    (\partial_g\omega)^2+(\partial_h\omega)^2
    }{
    4 \mathcal{J}^2\sin^2(\omega t)
    \left[
    4\langle \hat{n} \rangle(\langle \hat{n} \rangle+1)+2
    \right]
    },
    \quad
    t\to \infty ,
    \label{eq:qho_bound_opt}
\end{equation}
where the minimization has been performed over Gaussian states with a fixed average energy.

%Maximizing the variance of $\hat{Q}(t)$ is equivalent to maximizing the QFI of a squeezing unitary channel. For a capped mean photon number $\langle \hat{n}\rangle \leq \bar{n}$, the optimal variance is obtained by preparing a squeezed vacuum state~\cite{GAIBA2009,GenoniKerr2009,safranek2016}, $\ket{\psi_0} = \hat{S}(r_0 e^{i\theta_0})|0\rangle$ with $\theta_0 = \omega t$ and $\sinh^2r_0 = \bar{n}$. The resulting optimal variance is~\cite{GAIBA2009,safranek2016}
%\begin{equation}
%    \text{Var}_{\psi_0}(\hat Q(t)) = 2(2\bar{n}+1)^2.
%\end{equation}
%For a discussion of other constraints on Gaussian states, see Ref.~\cite{safranek2016}. Combining these results, we arrive at the formula
%\begin{equation}
%    \min_{\ket{\psi_0}} \Tr(F^{-1}) \simeq \frac{(\partial_g\omega)^2+(\partial_h\omega)^2}{2\sin^2(\omega t) (2\bar{n}+1)^2 \mathcal{J}^2}, \quad (t\to\infty).
%\end{equation}

%
%
\section{Example 3: the Lipkin--Meshkov--Glick model}
\label{sec:LMG}
We now apply the general framework to a system of $N$ qubits with a nonlinear interaction, focusing on the Lipkin--Meshkov--Glick (LMG) model in the strong-field regime. This provides a useful contrast to the previous examples: here the diagonal components of the Hamiltonian derivatives are linearly independent, so the obstruction responsible for the $t^0$ scaling does not arise. Nevertheless, the diagonal/off-diagonal decomposition can still shed light on the problem.

The Hamiltonian consists of a dominant rotation about $z$ and a nonlinear ``twisting'' interaction about $x$,
\begin{equation}
    \hat{H} = \omega \hat{J}_z - \frac{\lambda}{N} \hat{J}_x^2.
\end{equation}
\subsection{Diagonal/off-diagonal decomposition}
We assume $\omega \gg \lambda$, so that the eigenbasis of $\hat{H}$ is well-approximated by the Dicke states $\ket{j,m}$ (eigenstates of $\hat{J}_z$). In this regime, the commuting part of an operator is the diagonal part in the $z$ basis.

We seek to estimate $\boldsymbol{\theta}=(\omega,\lambda)$. The Hamiltonian derivatives are
\begin{align}
    \partial_\omega \hat{H} &= \hat{J}_z, \\
    \partial_\lambda \hat{H} &= -\frac{\hat{J}_x^2}{N}.
\end{align}
For $\omega$, the derivative is strictly diagonal in the Dicke basis,
\begin{equation}
    \hat{D}_\omega = \hat{J}_z, \qquad \hat{O}_\omega = 0.
\end{equation}
For $\lambda$, the decomposition yields
\begin{equation}
    \hat{D}_\lambda = \frac{1}{2N}\left(\hat{J}_z^2-\hat{\vect{J}}^2\right), \qquad \hat{O}_\lambda = -\frac{1}{4N}\left(\hat{J}_+^2 + \hat{J}_-^2\right).
\end{equation}
Since the diagonal generators are linearly independent (for any $N\ge 2$), the large-$t$ limit is dominated by the diagonal components,
\begin{equation}
    \hat{\mathcal{H}}_\omega \approx -t\,\hat{J}_z, 
    \qquad 
    \hat{\mathcal{H}}_\lambda \approx \frac{t}{2N}\left(\hat{\vect{J}}^2 - \hat{J}_z^2\right),
\end{equation}
with the off-diagonal contributions remaining bounded as $t\to\infty$. As a consequence, the QFIM acquires $t^2$ growth in both parameter directions, and quadratic time scaling in the scalar SLD bound is in principle achievable.

\subsection{Optimal initial state}

We restrict to the fully symmetric sector $j=N/2$, which maximizes the attainable variances of collective observables, and assume $N$ is even for simplicity. In this subspace $\hat{\vect{J}}^2=j(j+1)\,\hat{\mathbbm{1}}$ is fixed, so $\hat{D}_\lambda$ is equivalent (up to an irrelevant constant shift) to $\hat{J}_z^2/(2N)$. Thus, in the large-$t$ limit, the relevant diagonal generators are effectively $\hat{J}_z$ and $\hat{J}_z^2$. Our goal is to find the state that minimises the trace of the inverse of the covariance matrix of the diagonal generators.

The state that maximizes $\mathrm{Var}_{|\psi_0\rangle}(\hat{J}_z)$ is the GHZ state
\begin{equation}
    \ket{\mathrm{GHZ}}=\frac{1}{\sqrt{2}}\left(\ket{N/2}+\ket{-N/2}\right),
\end{equation}
whereas maximizing the spread of $\hat{J}_z^2$ favors placing weight near $m=0$ together with weight at the edges $m=\pm N/2$. Motivated by this structure, we consider the ansatz state
\begin{equation}
    \label{eq:LMGansatz}
    \ket{\psi}=\cos(\theta)\ket{0}+\frac{\sin(\theta)}{\sqrt{2}}\left(\ket{N/2}+\ket{-N/2}\right),
\end{equation}
which enforces equal amplitudes on $\ket{\pm N/2}$, thereby ensuring odd moments of $\hat{J}_z$ vanish while retaining a large second moment.

For this family, one finds
\begin{align}
    \mathrm{Var}_{|\psi_0\rangle}(\hat{J}_z) &= \frac{N^2}{4}\sin^2\theta,\\
    \mathrm{Var}_{|\psi_0\rangle}(\hat{J}_z^2) &= \frac{N^4}{16}\sin^2\theta\cos^2\theta,
\end{align}
and, since $\mathrm{Cov}(\hat{J}_z,\hat{J}_z^2)=0$ for the above symmetric superposition, the leading large-$t$ QFIM contributions are diagonal:
\begin{align}
    F_{\omega\omega} &= 4t^2\,\mathrm{Var}_{|\psi_0\rangle}(\hat{D}_\omega)=N^2t^2\sin^2\theta, \label{eq:Fomega}\\
    F_{\lambda\lambda} &= 4t^2\,\mathrm{Var}_{|\psi_0\rangle}(\hat{D}_\lambda)=\frac{N^2t^2}{16}\sin^2\theta\cos^2\theta. \label{eq:Flambda}
\end{align}
The SLD bound therefore reduces to
\begin{equation}
    \label{eq:LMGfulllimit}
    \mathrm{Tr}(WF^{-1})=\frac{w_\omega}{F_{\omega\omega}}+\frac{w_\lambda}{F_{\lambda\lambda}},
\end{equation}
where we have considered a diagonal weight matrix $W={\rm diag}(w_\omega, w_\lambda)$. As is apparent from Eqs.~\eqref{eq:Fomega} and~\eqref{eq:Flambda}, minimizing this over the angle $\theta$ defining the ansatz state \eqref{eq:LMGansatz} is equivalent to minimizing
\begin{equation}
    f(x)=\frac{w_\omega}{x}+\frac{16w_\lambda}{x(1-x)}, \qquad x:=\sin^2\theta\in[0,1].
\end{equation}
The minimum is attained at
\begin{equation}
    x=
    \frac{\sqrt{w_\omega+16w_\lambda}}
    {\sqrt{w_\omega+16w_\lambda}+4\sqrt{w_\lambda}},
\end{equation}
giving the SLD bound
\begin{equation}
    \mathrm{Tr}(WF^{-1}) = \frac{\left(\sqrt{w_\omega+16w_\lambda} + 4\sqrt{w_\lambda}\right)^2}{N^2t^2}.
\end{equation}

\subsection{Beyond the strong-field limit}
In the opposite $\lambda \gg \omega$ limit, the eigenbasis of $\hat{H}$ is well-approximated by the eigenstates of $\hat{J}_x$. When $N$ is even $\hat{J}_z$ is completely off-diagonal in this basis, causing $\hat{D}_\omega = 0$ and we revert to the $\mathcal{O}(t^0)$ scaling bottleneck for the SLD bound. Between these limits, we rely on numerics.
To sidestep the problem of choosing an initial state, we investigate the quantity $\Tr[G^{-1}]$ where $G$ is the Gram matrix of the diagonal generators defined in Eq.~\eqref{eq:Gram}. 
\begin{figure}[t]
\begin{center}
\includegraphics[angle=0,width=\columnwidth]{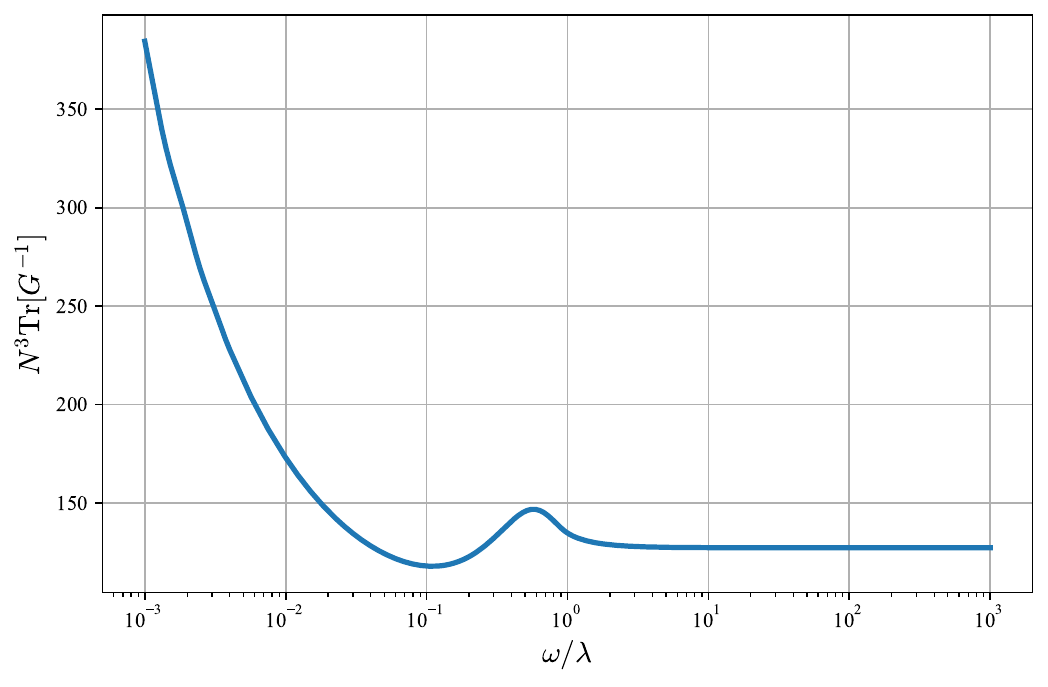} 
\end{center}
\caption{Trace of the inverse Gram matrix $\Tr[G^{-1}]$ of the {\it traceless diagonal generators} $\hat{\mathcal{D}_\mu}$ for the LMG model as a function of $\omega/\lambda$ with $N=128$.
}
\label{f:LMG}
\end{figure}
In Fig.~\ref{f:LMG} we plot $N^3\Tr[G^{-1}]$ as a function of $\omega/\lambda$ for $N=128$ and find that it changes very little from the strong-field regime to $\omega \approx \lambda$. In the weak driving limit, it diverges as expected due to the fact that, as discussed above, $\hat{D}_\omega \to 0$ in that limit. 

Additionally, we find that $\Tr[G^{-1}] \propto N^{-3}$ across all parameter regimes. 
By noticing that, up to a factor $d/(4 t^2)$, $\Tr[G^{-1}]$ is equivalent to the SLD bound evaluated for a probe state defined as an equally weighted superposition of the Hamiltonian eigenstates 
\begin{equation}
    \ket{\psi_0} = \frac{1}{\sqrt{N+1}} \sum_{k=1}^{N+1} \ket{E_k} ,
\end{equation}
this result confirms that the $N^{-2}t^{-2}$ scaling is possible for all parameters.
\section{Discussion}
\label{sec:discussion}
In the following, we will discuss our main results in several contexts: i) whether the $\mathcal{O}(t^0)$ bottleneck can be overcome by means of quantum control strategies; ii) how our results relate to measurement incompatibility and the more fundamental Holevo Cramér-Rao bound; and iii) the impact on our findings of more general weight matrices and the presence of nuisance parameters.
\subsection{Quantum Control Strategies}
In all of the examples considered so far, we have assumed that we are unable to add an additional control Hamiltonian. It is known that suitable control Hamiltonians can restore quantum-enhanced scaling of the QFI~\cite{HaidongPRL2016,pangOptimalAdaptiveControl2017}. The conceptually simplest and most effective control Hamiltonian is $\hat{H}_c = -\hat{H}(\mathbf{\tilde{g}})|_{\mathbf{\tilde{g}} = \mathbf{g}}$. This has the effect of setting the total Hamiltonian to zero, meaning that all Hamiltonian derivatives trivially commute with it. In this case, the local generators take the simple form
\begin{equation}
    \hat{\mathcal{H}}_\mu = -t (\partial_\mu \hat{H})
\end{equation}
and, in our scenario, the QFIM exhibits a clear $t^2$ scaling
\begin{equation}
    F_{\mu\nu} = 4t^2\,\mathrm{Cov}_{\ket{\psi_0}}(\partial_\mu \hat{H}, \partial_\nu \hat{H}). 
\end{equation}

We remark that the parameters $\mathbf{\tilde{g}}$ in the control Hamiltonian above are always constant (even when set equal to the true values $\mathbf{g}$); this is why the derivatives of the full Hamiltonian with respect to the parameters to be estimated are unaltered by the addition of the control. On the other hand, to restore the quadratic time scaling, one requires $\mathbf{\tilde{g}}=\mathbf{g}$, which demands a priori information. Therefore, in practice, this control must be applied adaptively~\cite{HaidongPRL2016,pangOptimalAdaptiveControl2017,wei2025}. In estimation problems with collective spin systems, such as those discussed in Sec.~\ref{sec:collectivespin}, such controls are implementable in practice and have been demonstrated experimentally~\cite{hou2021}. Conversely, the feasibility of applying these strategies to other scenarios, such as the one described in Sec.~\ref{sec:QHO}, may not be straightforward.

Although adaptive cancellation provides an optimal route to restore $t^{-2}$ scaling, it is not the only way. In fact, in certain cases, $t^{-2}$ scaling can be restored without any knowledge of the parameters to be estimated. For example, in the collective spin setting, adding a fixed nonlinear term such as $J_z^2$, or in the harmonic-oscillator setting, adding a fixed quartic term such as $\hat{x}^4$, takes the dynamics outside the original algebraic structure. Such additional terms lift the degeneracy in diagonal generators.
\subsection{Holevo Cram\'er-Rao bound and measurement incompatibility}
As we mentioned in Sec.~\ref{sec:multiparameter_qmetro}, the SLD bound $C_{S} = \Tr[F^{-1}]$ we introduced in Eq.~\eqref{eq:SLD_bound} is in general not saturable for simultaneous estimation of several parameters, because of the possible incompatibility of the optimal measurements identified by the SLD operators. The Holevo Cram\'er-Rao bound $C_H$ is a more fundamental bound, that is indeed always saturable for pure state families via separable measurements on an asymptotic number of copies $M$. We refer to~\cite{albarelli2020a} for more details on the definition and properties of the Holevo bound. The SLD and Holevo bounds obey the following chain of inequalities
\begin{equation}
    C_S \leq C_H \leq C_S + \big\| F^{-1}\mathcal U F^{-1}\big\|_1
    \label{eq:Holevo_chain}
\end{equation}
where we have assumed a weight matrix $W=\mathbbm{1}$, where $\|\cdot\|_1$ denotes the trace norm and $\mathcal U$ is the Uhlmann curvature with entries
\begin{equation}\label{eq:Uhlmann}
    \mathcal U_{\mu\nu}\;=\;\frac{1}{2i}\,\Tr\!\left(\hat{\rho}_\vect{g}\,[\hat{L}_\mu,\hat{L}_\nu]\right).
\end{equation}
The term $\big\| F^{-1}\mathcal U F^{-1}\big\|_1$ in the upper bound above can be interpreted as a measurement incompatibility penalty that is equal to zero for compatible parameters. For a unitary encoding of a pure probe state $\rho_0=\ketbra{\psi_0}$ one has
\begin{equation}
    \hat{L}_\mu \;=\; i\,2\,[\hat{\mathcal{H}}_\mu,\ketbra{\psi_0}],
\end{equation}
which allows us to express the Uhlmann curvature matrix elements directly in terms of the local generators:
\begin{equation}
    \mathcal{U}_{\mu\nu} = -2i \langle \psi_0 | [\hat{\mathcal{H}}_\mu, \hat{\mathcal{H}}_\nu] |\psi_0 \rangle.
\end{equation}
We are interested in the long-time limit ($t \to \infty$) of the incompatibility penalty $\|F^{-1}\mathcal{U}F^{-1}\|_1$ for a two-parameter estimation problem. Specifically, we analyze the scenario where the set of diagonal operators $\{\hat{D}_\mu\}$ is linearly dependent, which, as shown in Sec.~\ref{sec:scaling_proof}, causes the standard SLD bound to fall back to an $\mathcal{O}(t^0)$ scaling. 

Following the analysis performed in Secs.~\ref{sec:collectivespin} and~\ref{sec:QHO}, we derive the scaling of the incompatibility correction term, and we show that when the diagonal generators $\{\hat{D}_\mu\}$ are linearly dependent, the resulting ``slow'' direction in parameter space allows the Holevo bound to converge to the SLD bound as $\mathcal{O}(1/t)$.
Since Eq.~\eqref{eq:Holevo_chain} is invariant under an orthogonal reparameterization, we can rotate to the ``fast'' direction, $f$, and a ``slow'' direction, $s$. In this transformed basis, the local generators take the form:
\begin{align}
    \hat{\mathcal{H}}_f &= -t \hat{D}_f - \hat{K}_f(t), \quad (\text{where } \hat{D}_f \neq 0)\\
    \hat{\mathcal{H}}_s &= -\hat{K}_s(t), \quad (\text{since } \hat{D}_s = 0).
\end{align}
The determinant of the Fisher information matrix scales as $\mathcal{O}(t^2)$. 

Next, we evaluate the Uhlmann curvature in this basis. Since $\mathcal{U}$ is an antisymmetric matrix, the diagonal elements trivially vanish ($\mathcal{U}_{ff} = \mathcal{U}_{ss} = 0$). The off-diagonal element is given by
\begin{align}
    \mathcal{U}_{fs} &= -2i \langle [\hat{\mathcal{H}}_f, \hat{\mathcal{H}}_s] \rangle \nonumber \\
    &= -2i \langle [-t \hat{D}_f - \hat{K}_f, -\hat{K}_s] \rangle \nonumber \\
    &= 2i t \langle [\hat{D}_f, \hat{K}_s] \rangle + 2i \langle [\hat{K}_f, \hat{K}_s] \rangle.
\end{align}
Therefore, this matrix element grows at most linearly with time.

To find the incompatibility penalty, we compute the matrix $M = F^{-1}\mathcal{U}F^{-1}$. For any $2 \times 2$ symmetric matrix $F$ and antisymmetric matrix $\mathcal{U} = \begin{pmatrix} 0 & u \\ -u & 0 \end{pmatrix}$, the following identity holds:
\begin{equation}
    F^{-1}\mathcal{U}F^{-1} = \frac{u}{\det(F)} \begin{pmatrix} 0 & 1 \\ -1 & 0 \end{pmatrix}.
\end{equation}
The trace norm of this matrix is the sum of the absolute values of its singular values. The antisymmetric matrix on the right has singular values equal to one, yielding
\begin{equation}
    \|F^{-1}\mathcal{U}F^{-1}\|_1 = 2 \left| \frac{\mathcal{U}_{fs}}{\det(F)} \right|.
\end{equation}
Substituting our asymptotic scalings into this result gives:
\begin{equation}
    \|F^{-1}\mathcal{U}F^{-1}\|_1 = \frac{\mathcal{O}(t)}{\mathcal{O}(t^2)} = \mathcal{O}(t^{-1}).
\end{equation}

Thus, we have proven that when estimating two parameters with linearly dependent diagonal generators, the measurement incompatibility penalty decays as $1/t$. While the linear dependence of $\{\hat{D}_\mu\}$ prevents us from achieving quadratic time scaling ($\text{Tr}[F^{-1}] \sim \mathcal{O}(t^0)$), the vanishing curvature term guarantees that we do not suffer an additional asymptotic penalty from measurement incompatibility. In the long-time limit, the parameters can be estimated simultaneously and the Holevo bound converges to the SLD Cramér-Rao bound.
While the result above applies to the two-parameter case, we conjecture that for any number of parameters the $\mathcal{O}(t^0)$ contribution to the incompatibility is determined solely by the ``slow'' subspace.

We remark that in the example provided in Sec.~\ref{sec:LMG}, the diagonal generators commute and, therefore, the incompatibility vanishes asymptotically. This will always be the case for fast directions in non-degenerate Hamiltonians.
\subsection{Weight matrices and nuisance parameters}

Often, the multiparameter SLD bound (Eq.~\eqref{eq:SLD_bound}) is written with an additional positive semi-definite weight matrix $W$ to capture the relative importance of different parameters:
\begin{equation}
    \text{Tr}\left[W \text{Cov}(\check{\mathbf{g}})\right] \ge \text{Tr}\left(W F^{-1}\right).
    \label{eq:weighted_bound}
\end{equation}
Introducing this weight matrix prompts the question: Does reweighting the parameters alter the fundamental condition for the failure of quadratic time scaling? 

\subsubsection{Full rank weight matrices}
If $W$ is full rank, strictly positive definite and time-independent, it does not change the asymptotic analysis. A full-rank weight matrix can always be decomposed as $W = M^T M$ for some invertible matrix $M$. The weighted SLD bound can then be rewritten as:
\begin{equation}
    \text{Tr}(W F^{-1}) = \text{Tr}(M F^{-1} M^T) = \text{Tr}\left[ (M^{-T} F M^{-1})^{-1} \right].
\end{equation}
The matrix $F' = M^{-T} F M^{-1}$ is precisely the QFIM evaluated in a reparameterized basis $\mathbf{g}' = M \mathbf{g}$. Because the local generators transform linearly under a change of parameter basis ($\hat{\mathcal{H}}_{\mu}' = \sum_\nu (M^{-1})_{\mu \nu} \hat{\mathcal{H}}_\nu$), the linear dependence of the diagonal generators $\{\hat{D}_\mu\}$ is geometrically invariant. If the original set $\{\hat{D}_\mu\}$ is linearly dependent, the transformed set $\{\hat{D}_\mu'\}$ will also be linearly dependent, preserving the existence of a slow direction in the new parameter space. 

Furthermore, this coordinate transformation is already naturally embedded in the examples in Secs.~\ref{sec:collectivespin} and~\ref{sec:QHO}. In those examples, we allowed the Hamiltonian to have an arbitrary parameterization (e.g., $A(g,h)$ and $B(g,h)$ for the quantum harmonic oscillator), and demonstrated that the decomposition into fast and slow directions applies universally, regardless of the chosen parameter encoding. 

\subsubsection{Rank-deficient matrices and nuisance parameters}
The situation changes when $W$ is not full rank. The zero-eigenvalue subspace of $W$ corresponds to the presence of nuisance parameters, i.e., parameters (or linear combinations of parameters) that we do not know but that we do not care to estimate. As mentioned in Sec.~\ref{sec:scaling_proof}, the Gram matrix $G$ defined in Eq.~\eqref{eq:Gram} separates parameter space into directions with vanishing diagonal generators and directions with nonzero diagonal generators. To retain $\mathcal{O}(t^{-2})$ scaling the weight matrix must ignore all of the former directions. Therefore the mathematical condition for $\mathcal{O}(t^{-2})$ can be expressed concisely as
\begin{equation}
    \ker G \subseteq \ker W.
\end{equation}

Physically, this means that quadratic scaling is only recoverable if the slow parameter combinations are treated entirely as nuisance parameters.
To give further intuition, we consider a rank-1 weight matrix, $W = \mathbf{v} \mathbf{v}^T$, where $\mathbf{v}$ is a column vector. The weighted bound becomes:
\begin{equation}
    \text{Tr}(W F^{-1}) = \text{Tr}(\mathbf{v} \mathbf{v}^T F^{-1}) = \mathbf{v}^T F^{-1} \mathbf{v}.
\end{equation}

This represents the variance of estimating a single specific linear combination of parameters defined by $\mathbf{v}$. $F^{-1}$ contains $\mathcal{O}(t^0)$ along the slow directions and $\mathcal{O}(t^{-2})$ contributions along the fast directions. As a consequence, if $\mathbf{v}$ has any overlap with the slow subspace, the $\mathcal{O}(t^0)$ contribution will persist.

\section{Conclusions}
\label{sec:conclusions}
In this work, we established a general theoretical framework for understanding the fundamental limits of dynamical multiparameter quantum estimation under time-independent Hamiltonian dynamics. Specifically, we demonstrated that the failure to achieve simultaneous quadratic time scaling ($O(t^{-2})$) arises when the diagonal components of the Hamiltonian derivatives (those that commute with the Hamiltonian) are linearly dependent. This linear dependence creates a ``slow'' direction in parameter space where the variance is restricted to $O(t^0)$ scaling, bottlenecking the overall estimation precision. 

We illustrated the generality of this geometric obstruction across both discrete-variable and continuous-variable systems, including collective spin models and the generalized quantum harmonic oscillator. Although these models had clear reparameterizations into ``fast'' and ``slow'' parameters, the existence of such a reparameterization is not always so obvious~\cite{mihailescu2025a}. Conversely, we showed that when the diagonal generators remain linearly independent, as in the Lipkin–Meshkov–Glick model, simultaneous quadratic scaling is achievable and determined solely by the diagonal generators. Furthermore, our asymptotic analysis reveals that measurement incompatibility between slow and fast parameters decays as $1/t$, meaning the SLD bound is saturable in the long-time limit for all the examples considered. Ultimately, to overcome the $O(t^0)$ bottleneck in obstructed multiparameter problems, one must either relegate the slow directions to nuisance parameters via rank-deficient weight matrices or employ suitable quantum control techniques. 

A further practical implication of our analysis concerns the distinction between the interrogation time of a single coherent evolution and the total experimental time available. The obstruction identified here implies that, without additional control, the estimation precision remains bounded as $t \to \infty$. However, this does not imply that the slow direction is inestimable. One may instead reset the probe after a finite interrogation time $\tau$, or perform a sequence of measurements~\cite{burgarth2015quantum,radaelli2023fisher,o2024fisher,yang2025}. In this case, the Fisher information associated with the slow direction is $O(1)$ per shot, while the number of repetitions scales as $M \simeq T/\tau$ for a fixed total experimental time $T$. Optimizing over the finite interrogation time $\tau$, this strategy recovers $O(T^{-1})$ scaling of the variance, although it cannot restore the coherent $O(T^{-2})$ scaling without additional control. Thus, resetting or sequential measurements provide a useful practical compromise.

Our diagonal/off-diagonal decomposition therefore provides a clear, easily computable diagnostic tool for designing optimal multiparameter quantum metrology protocols and for deciding whether slow directions should be ignored, estimated through repeated measurements, or actively removed using adaptive control.

\section*{Acknowledgments}
EOC and MGG acknowledge discussions with Francesco Albarelli.

%
%\end{document}

\appendix
\section{Derivation of the local generator}
\label{app:generator_collective}

For any parameter $g$, the local generator can be written in integral form as
\begin{equation}
\hat{\mathcal{H}}_g \;=\; i(\partial_g \hat{U}^\dagger)\,\hat{U}
\;=\; -\!\int_{0}^{t}\! ds\; e^{is \hat{H}}\,(\partial_g \hat{H})\,e^{-is\hat{H}}.
\label{eq:integral_rep}
\end{equation}
Due to the relationship between SU(2) and SO(3), we can write this Heisenberg picture evolution in terms of the 3D rotation matrix around $\uvect{n}$
\begin{equation}
e^{is\hat{H}}\,(\mathbf a\cdot\hat{\vect J})\,e^{-is\hat{H}}
\;=\; \big(R_{\uvect{n}}(-\Omega s)\,\mathbf a\big)\cdot\hat{\vect J}
\label{eq:adjoint_action}
\end{equation}
where $R_{\uvect{n}}(\theta)$ denotes a three-dimensional rotation by angle $\theta$ around the axis ${\uvect{n}}$.

Since $\partial_g \hat{H} = (\partial_g \Omega)\,\uvect{n}\cdot\hat{\vect J} + \Omega\,(\partial_g\uvect{n})\cdot\hat{\vect J}$,
Eq.~\eqref{eq:integral_rep} becomes
\begin{equation}
\hat{\mathcal H}_g = -(\partial_g \Omega)\!\int_0^t\!ds\;
e^{is\hat{H}}(\uvect{n}\cdot\hat{\vect J})e^{-is\hat{H}}
\;-\;
\Omega\!\int_0^t\!ds\;
e^{is\hat{H}}(\partial_g\uvect{n}\cdot\hat{\vect J})e^{-is\hat{H}}.
\label{eq:split}
\end{equation}

The first term is straightforward, because $\uvect{n}\cdot\hat{\vect J}$ commutes with $\hat{H}$:
\begin{equation}
-(\partial_g\Omega)\!\int_0^t\!ds\,\uvect{n}\cdot\hat{\vect J}
= -t(\partial_g\Omega)\,\uvect{n}\cdot\hat{\vect J}.
\label{eq:first_term}
\end{equation}

For the second term, we use that $\partial_g\uvect{n}\perp\uvect{n}$, so under the rotation
$R_{\uvect{n}}(-\Omega s)$ we have
\begin{equation}
R_{\uvect{n}}(-\Omega s)\,\partial_g\uvect{n}
= \partial_g\uvect{n}\cos(\Omega s)
- (\uvect{n}\times\partial_g\uvect{n})\sin(\Omega s).
\label{eq:rotation_on_dndg}
\end{equation}
Substituting into Eq.~\eqref{eq:split} gives
\begin{align}
\hat{\mathcal H}_g^{(\perp)} &= -\Omega\!\int_0^t\!ds\;
e^{is\hat{H}}(\partial_g\uvect{n}\cdot\hat{\vect J})e^{-is\hat{H}} \nonumber\\ 
&= -\Omega\!\int_0^t\!ds\;
\bigg(\big[ \partial_g\uvect{n}\cos(\Omega s) 
 - (\uvect{n}\times\partial_g\uvect{n})\sin(\Omega s)\big]\cdot\hat{\vect J} \,\bigg) \nonumber\\
&= -\Big[ \sin(\Omega t)\,(\partial_g\uvect{n}) - (1-\cos(\Omega t))\,(\uvect{n}\times\partial_g\uvect{n}) \Big]\cdot\hat{\vect J},
\label{eq:second_term}
\end{align}
where we used the integral identities
\begin{align}
\int_0^t\!\!ds\,\cos(\Omega s)&=\frac{\sin(\Omega t)}{\Omega}\\
\int_0^t\!\!ds\,\sin(\Omega s)&=\frac{1-\cos(\Omega t)}{\Omega}.
\end{align}
Combining Eqs.~\eqref{eq:first_term} and \eqref{eq:second_term}, we find that the local generator has the compact vector form
\begin{align}
\hat{\mathcal H}_g &= \mathbf h_g\cdot\hat{\vect J}, \\
\mathbf h_g &= -t(\partial_g\Omega)\,\uvect{n}
- \sin(\Omega t) \,\partial_g\uvect{n}
+ (1-\cos(\Omega t))\,(\uvect{n}\times\partial_g\uvect{n}).
\label{eq:h_vector_final}
\end{align}
This represents an alternative derivation of Eq.~(12) of Ref.~\cite{jing2015} and Eq.~(22) of Ref.~\cite{yang2022multiparameter}.

\section{Optimal Gaussian initial state for the quantum harmonic oscillator estimation problem}
\label{app:qho_min_eig}

In this appendix, we derive the optimal initial state for the example presented in Sec.~\ref{sec:QHO}. As discussed in the manuscript, the goal is to maximize the slow eigenvalue
\begin{equation}
    \lambda_\text{min} 
    \approx 
    \frac{4 \mathcal{J}^2\sin^2(\omega t)}{\|\vect{w}\|^2}
    \left( 
    \mathrm{Var}\left(\hat{Q}(t)\right)  
    - 
    \frac{\mathrm{Cov}\left(\hat{n}, \hat{Q}(t)\right)^2}{\mathrm{Var}\left(\hat{n} \right)} 
    \right),
    \label{eq:qho_lmin_app}
\end{equation}
over pure Gaussian states $\ket{\psi_0}$ with fixed average bosonic excitations $\langle \hat{n} \rangle$. All expectation values, variances, and covariances are taken with respect to $\ket{\psi_0}$.

It is convenient to remove the explicit phase in $\hat Q(t)$ by introducing the rotated mode $\hat b=e^{-i\omega t/2}\hat c$, so that
\begin{equation}
    \hat Q(t)=\hat b^2+\hat b^{\dagger 2},
    \qquad
    \hat n=\hat b^\dagger \hat b .
\end{equation}
The number operator is invariant under this phase rotation. The most general pure single-mode Gaussian state can be written as
\begin{equation}
    |\psi_0\rangle
    =
    \hat D_b(\alpha)\hat S_b(s e^{i\theta})|0\rangle .
\end{equation}
We denote
\begin{equation}
    N_s=\sinh^2 s, \qquad
    N_d=|\alpha|^2, \qquad
    \langle \hat{n} \rangle=N_s+N_d ,
    \label{eq:displacement_squeezing}
\end{equation}
where $N_s$ is the number of photons in the squeezed vacuum and $N_d$ is the number of photons in the displacement. We also define the additional parameter
\begin{equation}
    M=\sqrt{N_s(N_s+1)} .
\end{equation}

\par
For any two-dimensional matrix 
\begin{equation}
    \Sigma=
    \begin{pmatrix}
        a & \gamma \\
        \gamma & b
    \end{pmatrix},
\label{eq:twod_matrix}
\end{equation}
we define the Schur complement
\begin{equation}
    S(\Sigma)
    =
    b-\frac{\gamma^2}{a}.
\end{equation}
By inspecting the formula for the minimum eigenvalue in Eq.~\eqref{eq:qho_lmin_app}, it is then evident that its maximization corresponds to the maximization of the Schur complement of the covariance matrix for the operators $\hat n$ and $\hat{Q}(t)$, 
\begin{align}
\Sigma =  
\begin{pmatrix}
    \mathrm{Var}\left(\hat{n} \right) & \mathrm{Cov}\left(\hat{n}, \hat{Q}(t)\right) \\
    \mathrm{Cov}\left(\hat{n}, \hat{Q}(t)\right) & \mathrm{Var}\left(\hat{Q}(t)\right)
\end{pmatrix}.
\end{align}
We remark that this result is not peculiar to this scenario: in the example discussed in Sec.~\ref{sec:collectivespin}, the goal was indeed to maximize the Schur complement of the covariance matrix for the pair of operators $(\uvect n \cdot \hat{\vect{J}}, \vect{h}_{\vect{w}} \cdot \hat{\vect{J}})$ corresponding to fast and slow directions in the parameter space.

For any operator $\hat{x}$ we define the centered operator as
\begin{equation}
    \Delta \hat{x} = \hat{x} - \langle \hat{x} \rangle.
\end{equation}
Using $\langle \hat{b} \rangle = \alpha$ we have
\begin{align}
    \Delta \hat n
    &=
    \left(\Delta \hat b^\dagger \Delta \hat b-N_s\right)
    +
    \left(\alpha^\ast\Delta \hat b+\alpha \Delta \hat b^\dagger\right),
    \\
    \Delta \hat Q
    &=
    \left[
    \Delta \hat b^2+\Delta \hat b^{\dagger 2}
    -
    \left\langle
    \Delta \hat b^2+\Delta \hat b^{\dagger 2}
    \right\rangle
    \right]
    +
    2\left(\alpha\Delta \hat b+\alpha^\ast \Delta \hat b^\dagger\right).
\end{align}
These terms decompose into centered squeezed contribution (on the left) and the displacement contributions (on the right). Under our Gaussian states assumptions, we find that all mixed covariances between the centered quadratic terms and the linear displacement terms vanish, because they contain odd centered Gaussian moments. Therefore, the covariance matrix of the pair $(\hat n,\hat Q)$ decomposes into squeezed and displaced contributions:
\begin{equation}
    \Sigma=\Sigma_s+\Sigma_d.
\end{equation}

We now proceed to evaluate the corresponding matrix elements. We remind the fact that, since the state is Gaussian, all moments are eventually determined by the first and second moments of $\hat{b}$ and $\hat{b}^\dag$.

We first consider the centered squeezed contribution $\Sigma_s$. Using $\langle \Delta \hat b^2\rangle = -M e^{i\theta},$ $\langle \Delta \hat b^\dag\Delta \hat b\rangle = N_s,$ and Wick's theorem, the corresponding matrix entries, using the convention introduced in Eq.~\eqref{eq:twod_matrix}, are
\begin{align}
    a_s&=2M^2,
    \\
    b_s&=2+8M^2\cos^2\theta,
    \\
    \gamma_s&=-2(2N_s+1)M\cos\theta .
\end{align}
We now consider the displacement contribution. Defining $\alpha = \sqrt{N_d} e^{i\phi}$, the corresponding matrix entries are
\begin{align}
    a_d
    &=
    N_d
    \left[
    (2N_s+1)-2M\cos(\theta-2\phi)
    \right],
    \\
    b_d
    &=
    4N_d
    \left[
    (2N_s+1)-2M\cos(\theta+2\phi)
    \right],
    \\
    \gamma_d
    &=
    2N_d
    \left[
    (2N_s+1)\cos(2\phi)-2M\cos\theta
    \right].
\end{align}
A direct calculation of the Schur complement gives
\begin{equation}
    S(\Sigma_s+\Sigma_d)
    =
    S(\Sigma_s)+S(\Sigma_d)
    -
    \frac{
    (a_d\gamma_s-a_s\gamma_d)^2
    }{
    a_s a_d(a_s+a_d)
    }.
    \label{eq:appB_schur_add}
\end{equation}
Since they are variances, $a_s$ and $a_d$ are non-negative and thus
\begin{equation}
    S(\Sigma_s+\Sigma_d)
    \le
    S(\Sigma_s)+S(\Sigma_d).
    \label{eq:appB_schur_bound}
\end{equation}
We now proceed in trying to maximize simultaneously both Schur complements $ S(\Sigma_s)$ and $ S(\Sigma_d)$. 
Since $(2N_s+1)^2=4M^2+1$, one obtains
\begin{equation}
    S(\Sigma_s)
    =
    2\sin^2\theta
    \le 2 .
    \label{eq:appB_s_bound}
\end{equation}
As regards $\Sigma_d$, we can simultaneously maximize $b_d$ and set $\gamma_d = 0$ using $\cos(\theta+2\phi) = -1$, $\cos(2\phi) = 0$ and $\cos\theta = 0$. Conveniently, this also maximizes Eq.~\eqref{eq:appB_s_bound} while keeping $\gamma_s=0$. Since both $\gamma_s$ and $\gamma_d$ are zero, we do also saturate Eq.~\eqref{eq:appB_schur_bound}. The result is
\begin{equation}
    S(\Sigma)
    =
    2+4N_d(2N_s+1+2M),
    \label{eq:appB_phase_bound}
\end{equation}
with some example values for saturation
\begin{equation}
    \theta = \frac{\pi}{2},
    \quad
    \phi = \frac{\pi}{4}.
\end{equation}
These phases are expressed in terms of the squeezing and displacement operators of the rotated mode $\hat{b}$. To express the same physical state in terms of the original mode $\hat c$, we use
\begin{equation}
    \hat D_b(\alpha)=\hat D_c\!\left(\alpha e^{i\omega t/2}\right),
    \qquad
    \hat S_b(se^{i\theta})=\hat S_c\!\left(se^{i(\theta+\omega t)}\right),
\end{equation}
Therefore, the optimal phases are
\begin{equation}
    \phi_c=\frac{\pi+2\omega t}{4},
    \qquad
    \theta_c=\frac{\pi+2\omega t}{2}.
\end{equation}

What's left is optimizing the energy split between squeezing and displacement. Substituting $N_d=\langle \hat{n}\rangle-N_s$ into Eq.~\eqref{eq:appB_phase_bound}, we need to maximize
\begin{equation}
    f(N_s)
    =
    (\langle \hat{n}\rangle-N_s)
    \left(
    2N_s+1+2\sqrt{N_s(N_s+1)}
    \right).
\end{equation}
There is a unique stationary point at
\begin{equation}
    N_s^\star
    =
    \frac{\langle \hat{n}\rangle^2}{2\langle \hat{n}\rangle+1},
    \qquad
    N_d^\star
    =
    \frac{\langle \hat{n}\rangle(\langle \hat{n}\rangle+1)}{2\langle \hat{n}\rangle+1},
    \label{eq:appB_opt_split}
\end{equation}
which thus fix the optimal values of the displacement and squeezing parameters $|\alpha|$ and $s$ via Eqs.~\eqref{eq:displacement_squeezing}.

Combining all the above results, we finally obtain
\begin{equation}
    \lambda_\text{min}^{\mathrm{opt}}
    \approx
    \frac{4 \mathcal{J}^2\sin^2(\omega t)}{\|\vect{w}\|^2}
    \left[
    4\langle \hat{n} \rangle(\langle \hat{n} \rangle+1)+2
    \right].
    \label{eq:appB_lmin_opt}
\end{equation}
\bibliography{refs}
\end{document}